\documentclass{sn-jnl}

\usepackage{algorithm}
\usepackage{algorithmicx}
\usepackage{algpseudocode}
\usepackage[title]{appendix}
\usepackage{amsfonts}
\usepackage{amsmath}
\usepackage{amssymb}
\usepackage{amsthm}
\usepackage{bm}
\usepackage{booktabs}
\usepackage{comment}
\usepackage{dsfont}
\usepackage{enumerate}
\usepackage{graphicx}
\usepackage{listings}
\usepackage{mathrsfs}
\usepackage{mathtools}	
\usepackage{manyfoot}
\usepackage{multirow}
\usepackage{physics}
\usepackage{soul}
\usepackage{siunitx}
\usepackage{textcomp}
\usepackage[dvipsnames]{xcolor}

\usepackage{setspace}
\usepackage{caption}
\usepackage{threeparttable}
\newcommand{\oa}{\mathcal{O}_{a}}
\newcommand{\ob}{\mathcal{O}_{b}}

\newcommand{\ka}{K_{a}}
\newcommand{\kb}{K_{b}}

\newcommand{\pl}{\mathcal{P}_{l}}
\newcommand{\pr}{\mathcal{P}_{r}}

\newcommand{\uu}{\vb{u}}
\newcommand{\pp}{\vb{P}}
\newcommand{\ep}{E_{p}}

\newcommand{\eb}{\mathscr{E}}
\newcommand{\mb}{\mathscr{P}}

\newcommand{\p}[1]{\partial_{#1}}

\newcommand{\RR}{\mathbb{R}}
\newcommand{\ZZ}{\mathbb{Z}}

\theoremstyle{remark}
\newtheorem{remark}{Remark}

\begin{document}
\author{\fnm{Tony} \sur{Lyons}}\email{tony.lyons@setu.ie}
\affil{\orgdiv{Department of Computing and Mathematics}, \orgname{South East Technological University}, \orgaddress{\street{Cork Road}, \city{Waterford}, \country{Ireland}}}

\title{Relational space-time and de Broglie waves}

\maketitle

\abstract{
Relative motion of particles is examined in the context of relational space-time. It is shown that de Broglie waves may be derived as a representation of the coordinate maps between the rest-frames of these particles. Energy and momentum are not absolute characteristics of these particles, they are understood as parameters of the coordinate maps between their rest-frames. It is also demonstrated the position of a particle is not an absolute, it is contingent on the frame of reference used to observe the particle.
}

\section{Introduction}
\subsection{Relational space-time}\label{sec:relational}
In this paper we consider the relative motion of material point particles in the context of relational space-time and aim to show that de Broglie waves\footnote{de Broglie waves as defined by Dirac \cite{Dir2007} p.120} may be deduced as a representation of these point particles.
In \cite{Bar1982} Barbour examines in detail the development of relational concepts of space and time  from Leibniz \cite{LC2000} up to and including his own work on relational formulations of dynamics  \cite{Bar1974,BB1977,BB1982}. A central point of discussion in \cite{Bar1982} is that the uniformity of space means its points are indiscernible, which are made discernible only by the presence of ``substance.''\footnote{In the sense used by Minkowski, Cologne (1908) \cite{Min2012}} This relational understanding of space and time supposes it is the varied and changing distribution of matter which endows space-time with enough variety to distinguish points therein.

In \cite{Mun1983}, Mundy describes absolutist theories of matter, space and time as those which treat space-time points as entities in their own right.  In contrast, relational theories assume only physical objects as the basic entities, the properties of space and time being deduced from the relations among these physical objects (see also \cite{Skl1974}). In the relational framework, only physical objects are primary, while space and time coordinates are realised in terms of the changing relative configurations of these objects. An extensive survey of relational interpretations of post-Newtonian space, time and motion may be found at \cite{HHR2023}, while comprehensive account of the philosophical development of space, time and mechanics is presented in \cite{Hug1999}.

\begin{figure}[!ht]
\centering
\captionsetup{width=\textwidth}
\includegraphics[width=\textwidth]{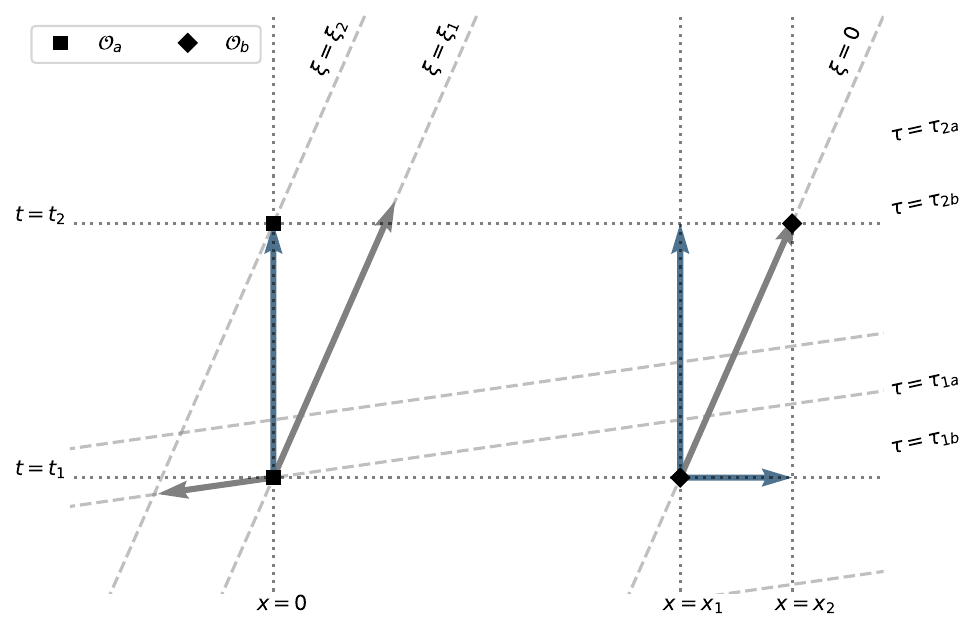}
\caption{The relative motion of $\oa$ and $\ob$ and the coordinate displacements this defines in the reference frames $K_{a}$ and $K_{b}$.}\label{fig:motionab}
\end{figure}
Figure \ref{fig:motionab} illustrates point-like observers $\oa$ and $\ob$  with associated rest-frames $\ka$ and $\kb$, in a state of relative motion. In the frame $\ka$ it appears the observer $\ob$ moves between space-time locations $(t_{1},x_{1})$ and $(t_{2},x_{2})$ , while $\oa$ ``moves'' between locations $(t_{1},0)$ and $(t_{2},0)$. The point particle $\oa$ \emph{defines} the location $x=0$ in the reference frame $\ka$.
In the simplest case, the initial distance between $\oa$ and $\ob$ with reference to $\ka$ is $\abs{x_{1}-0}$ and so we say $\ob$ occupies $x_{1}$ if the distance between $\oa$ and $\ob$ is $\abs{x_{1}-0}$. That is to say, the coordinate $x_{1}$ is defined as $\pm\abs{x_{1}-0}$ for $x_{1}\geq0$ or $x_{1}<0$ respectively. In the frame $K_{b}$ the observer $\ob$ is seen to ``move''  between space-time locations of the form $(\tau_{1b},0)$ and $(\tau_{2b},0)$ while $\oa$ moves between $(\tau_{1a},\xi_{1})$ and $(\tau_{2a},\xi_{2})$.  The spatial separation between the points $(t_{1},x_{1})$ and  $(t_{2},x_{2})$ is simply not recognised in the rest frame of $\ob$ in the relational framework. On the contrary, the locations $x=x_{1}$ and $x=x_{2}$ are made discernible only because the material point $\ob$ is observed to move between these locations.

Similarly, the instants $t=t_{1}$ and $t=t_{2}$ are made discernible only by the changing location of $\ob$ with respect to $\oa$, in line with Barbour's Leibnizian view of time as successive configurations of material bodies (see \cite{Bar1974, LC2000}). Indeed it is such material re-configurations which allow for the measurement of time intervals in practice. For instance, the motion of a sprinter between two fixed positions on a race-track is compared to the number of periodic vibrations of a quartz crystal, typically oscillating at $2^{15}$ Hz in modern watches.  The relational viewpoint suggests that the instants $t=t_{1}$ and $t=t_{2}$ have no intrinsic separation (or indeed meaning)  without reference to the observed motion of $\ob$ between the locations $x=x_{1}$ and $x=x_{2}$.

In this regard,  the Machian formulation of dynamics developed by Barbour \& Bertotti in \cite{BB1982} is of particular interest. The Machian dynamics developed by these authors considered a system of $N$ particles whose configuration space is $Q\owns q=(\bm{r}_{1},\ldots,\bm{r}_{N})$, where $\left\{\bm{r}_{n}\in\RR^{3}\right\}_{n=1}^{N}$ are the locations of the individual particles. A history of this system of particles is simply a parameterised path $q(\lambda)\in Q$, where $\lambda$ is a monotonic and continuous parameter which labels each configuration of particles along the history of the system. It is supposed that this path is governed by an action $S=\int \mathscr{L}[q,q']\dd\lambda$. This dynamical theory is said to implement Mach's first and second principles if it is invariant under the action of Euclidean group: $\bm{r}(\lambda)\to \vb{R}(\lambda)\cdot\bm{r}(\lambda)+\bm{h}(\lambda)$ where $\left(\vb{R}(\lambda),\bm{h}(\lambda)\right)\in O(3)\ltimes \RR^3$ (Mach's $\mathrm{1^{st}}$ principle) and arbitrary re-parameterisations $\lambda\to f(\lambda)$ such that $f'(\lambda)\neq0$ (Mach's $\mathrm{2^{nd}}$ principle).

In the same work, Barbour \& Bertotti constructed a metric on the space of orbits  $Q_{0}\owns\left\{q\right\}$ which are the subsets of $Q$ invariant under the action of the Euclidean group. This metric was in turn used to deduce an intrinsic time variable from the separation between two elements in $Q_{0}$, whereby they were able to recover classical Newtonian mechanics in a Machian sense. It was also demonstrated that their framework is consistent with Lorentz invariant field theory and more generally gauge theories and general relativity are examples of the intrinsic time dynamics they construct. A comprehensive survey of this and other contemporary  formulations of Machian dynamics is  presented in \cite{Mer2018}.

In line with the work of Barbour \& Bertotti, in this paper time intervals are also understood to manifest through changing configurations of material objects. In contrast however, it will be argued that time variables emerges as an intrinsic property of the coordinate maps between the rest frames of the point-like observers. It is understood that the \emph{coordinate differences} in each frame of reference serve to characterise the relative motion. For instance it is the coordinate difference $(t_{2}-t_{1},x_{2}-x_{1})$ which serve to define the velocity and related energy-momentum of $\ob$ with reference to $K_a$. Here, the  \emph{mappings of coordinate differences} between reference frames provides the framework to describe the relative motion of the observers $\oa$ and $\ob$, while these coordinate maps themselves are found to be naturally represented via de Broglie waves.

\subsection{Relativity and de Broglie waves}\label{sec:debroglie}
It is assumed the observer $\ob$ moves with reference to $K_{a}$ at constant velocity $v=\beta c$, where $\beta\in(-1,1)$ and $c$ is the speed of light.
In the relational context, this obviously means the change of spatial distance between $\oa$ and $\ob$ from $\abs{x_{b}}$ to $\abs{x_{b}+\beta c\dd{t}}$ serves to define the time interval $\dd{t}$.
The coordinate map $\bm{\Xi}:K_{a}\to K_{b}$ takes the form
\begin{equation}\label{eq:Lorentz}
\tau=\gamma\left(t-\frac{\beta}{c}x\right)\quad \xi = \gamma\left(x-c\beta t\right);\quad \gamma=\frac{1}{\sqrt{1-\beta^2}}.
\end{equation}
\begin{remark}
The coordinate map \eqref{eq:Lorentz} is more accurately understood as a map of the coordinate differences $(t-0,x-0)$ to $(\tau-0,\xi-0)$. If one generalises to the case where some arbitrary coordinate  $(t_{0},x_{0})$ coincide with $(\tau_{0},\xi_{0})$, then following Einstein's derivation \cite{Ein1905} yields the  map
\begin{equation}\label{eq:Lorentz-gen}
\begin{aligned}
\tau-\tau_{0} &=\gamma\left((t-t_{0})-\frac{\beta}{c}(x-x_{0})\right)\\
\xi-\xi_{0} &= \gamma\left((x-x_0)-c\beta(t-t_{0})\right).
\end{aligned}
\end{equation}
\end{remark}
The point emphasised by de Broglie \cite{deB1925,deB1930} is $\ob$ has an associated angular frequency
\begin{equation}
 \omega_{0} = \frac{E_{0}}{\hbar} ,
\end{equation}
which may be obtained from the Planck and Einstein relations $E=\hbar\omega$ and $E_0=mc^2$, where $m$ is the rest mass of $\ob$.
In \cite{Gou2005, Cat2008} the authors describe a novel set of experiments using channeling motion of \SI{80}{\mega\electronvolt} electrons  undergoing rosette motion  upon interaction with a single layer of atoms in a silicon crystal. It is argued by the authors that these experiments may be interpreted as a direct observation of the so-called internal clock of the electron.

Given this intrinsic angular frequency $\omega_{0}$, de Broglie postulated that the wave-form $\psi(\tau,\xi) = e^{i\omega_{0}\tau}$ is naturally associated with the observer $\ob$. Meanwhile \eqref{eq:Lorentz} ensures this wave-form with respect to $K_{a}$ is of the form
\begin{equation}\label{eq:wf}
\psi(t,x) = e^{i\omega_{0}\gamma\left(t-\frac{\beta}{c}x\right)} = e^{i(\omega t-kx)},
\end{equation}
where $\omega = \gamma\omega_{0}$ and $k=\frac{\omega_{0}\beta\gamma}{c}=\frac{\beta}{c}\omega$. The relativistic energy and momentum of $\ob$ with reference to $K_{a}$ are given by $E= mc^2\gamma$ and $p=mc\beta\gamma$, and as such the wave-form $\psi(t,x)$ may be also written as
\begin{equation}\label{eq:deB1}
\psi(t,x) = e^{i(\omega t-kx)} := e^{\frac{i}{\hbar}(Et-px)}.
\end{equation}
Thus the relativistic energy-momentum $(E,p)$ of the observer $\ob$ are related to the angular frequency $\omega$ and wave-number $k$ of the associated wave-form $\psi$.

\begin{figure}[!ht]
\centering
\captionsetup{width=\textwidth}
\includegraphics[width=\textwidth]{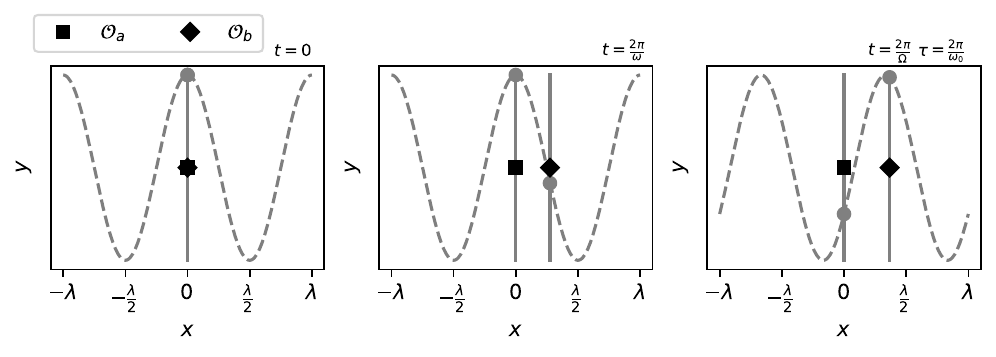}
\caption{Snapshots of the relative motion of $\oa$ and $\ob$, their local clocks with frequencies $\omega$ and $\omega_{0}$ and the wave-form $\psi(t,x) = \cos(\omega t - kx)$. A related animation may be found at: \href{https://link.springer.com/article/10.1007/s10701-023-00715-9}{de Broglie wave animation}}\label{fig:waveform}
\end{figure}

A point of importance for de Broglie was that the wave form $\psi(t,x)$ is always in phase with a clock of period $T_{0}=\frac{2\pi}{\omega_{0}}=\frac{mc^2}{\hbar}$ at rest in the frame $K_{b}$
(see \cite{Loc1984} for an interesting account of the significance de Broglie accorded to this result).
This clock is shown in Figure \ref{fig:waveform} as an oscillator moving along the $y$-axis of the frame $K_{b}$ with angular frequency $\omega_{0}$.
The period and angular frequency of this clock relative to $K_{a}$ are
\begin{equation}
 T = \gamma T_{0}\quad  \Omega = \frac{2\pi}{T} = \frac{\omega_{0}}{\gamma}.
\end{equation}
The angular frequency $\Omega$ is not to be confused with the angular frequency of $\psi(t,x)$ which is $\omega=\gamma\omega_{0}$. For reference Figure \ref{fig:waveform} also shows a similar clock at rest in $K_{a}$ with angular frequency $\omega$ (see \cite{McK1976} for further discussion of the interrelationship of the angular frequencies $\omega_{0}$, $\omega$ and $\Omega$).

The clock co-moving with $\ob$ between $(t,x)$ and $(t+\dd t,x+\beta c\dd t)$ in $\ka$ will undergo a phase-shift $\dd\Phi=\Omega \dd t = \frac{\omega_{0}}{\gamma}\dd t$.
Meanwhile, the phase difference of the wave $\psi(t,x)$, between $(t,x)$ and $(t+\dd t,x+\beta cdt)$  is
\begin{equation}
\omega_{0}\gamma\left(\dd t - \frac{\beta}{c}\beta c\dd t\right) =\frac{\omega_0}{\gamma}\dd t= \dd\Phi,
\end{equation}
so the moving clock and wave-form $\psi(t,x)$ are in phase, (see Figure \ref{fig:waveform} and the associated animation). It is clear then that de Broglie waves are closely connected with the Lorentz transformation between local inertial reference frames $K_{a}$ and $K_{b}$, in particular with the coordinate map $\tau(t,x)$. The existence of de Broglie waves was confirmed almost immediately after de Broglie's first prediction \cite{deB1925}, with the interference experiments of Davisson \& Germer \cite{DG1927} and  the contemporaneous experiments of Thomson \& Reid \cite{TR1927}. In the years since, the experimental evidence supporting de Broglie's conjecture has accumulated steadily (see \cite{Arnetal1999,Setal2008,Setal2020} among others). A comprehensive treatment of the wave-mechanics of de Broglie waves as an extension of Hamilton's optical methods in the context of absolute space-time is given in \cite{Syn1954}.

The emergence of Minkowski space time from de Broglie waves is explored in \cite{Kas2013}, where it was argued by Kastner that de Broglie waves, despite their super-luminal phase speed must be accorded some physical significance for two reasons: Firstly, the wave length of a de Broglie wave associated with a particle is in fact measured routinely in diffraction experiments. Secondly, de Broglie waves may actually provide a dynamical structure from which Minkowski space time emerges in a relational framework. It is argued that given the infinite phase speed of a de Broglie wave in the rest frame of the associated physical object, then the phase aspect of the de Broglie wave essentially serves as the spatial axis of this rest frame. Moreover, given that the group velocity associated with the wave coincides with the actual velocity of this object in any frame, then the group aspect of this de Broglie wave may serve as a time axis for the rest frame of the object.

The work to follow here is in apparent agreement with Kastner's proposal. However, our aim here is to deduce from first principles why de Broglie waves must appear as a representation of the coordinate maps between the rest-frames of the observers $\oa$ and $\ob$, in the framework of relational space-time.

\section{Coordinate maps and their governing equations}
\subsection{Coordinate maps \& relativity}\label{sec:existence}
At the simplest level, the approach we adopt here follows from the observation that $\ob$ moves through $\ka$ with velocity $\beta c$ if and only if $\oa$ moves through $\kb$ with velocity $-\beta c$ (assuming the space-time axes of $\ka$ and $\kb$ have the same orientation).  With reference to Figure \ref{fig:motionab}, this means the coordinates of $\ka$ and $\kb$ must be identified according to
\begin{align*}
(t_{1},0)\sim (\tau_{1a},\xi_{1}) && (t_{2},0)\sim (\tau_{2a},\xi_{2}) \\
(t_{1},x_{1})\sim (\tau_{1b},0) && (t_{2},x_{2})\sim (\tau_{2b},0),
\end{align*}
and more importantly
$$
\frac{x_{2}-x_{1}}{t_{2}-t_{1}} = -\frac{\xi_{2}-\xi_{1}}{\tau_{2a}-\tau_{1a}}=\beta c.
$$
That is to say, to ensure the relative motion is physical, it is necessary that the coordinate differences $(t_{2}-t_{1},x_{2}-x_{1})$ and $(\tau_{2}-\tau_{1},\xi_{2}-\xi_{1})$ transform among themselves in a consistent way.

In line with \S \ref{sec:relational}, we note that the coordinates $(t_{1},x_{1})$ only have meaning due to $\ob$ being located a distance $\abs{x_{1}}$ from $\oa$, with this configuration of observers defining the instant $t_{1}$. That is to say, the point $(x_{1},t_{1})$ is physically realised since it corresponds to the location $(\tau_{1b},0)$. Similarly, the location $(t_{2},x_{2})\in \ka$ has physical significance only because it is occupied by $\ob$ or in other words because it corresponds to the location $(\tau_{2b},0)\in \kb$. Conversely, it is also clear that the coordinates $(\tau_{1a},\xi_{1})$ and $(\tau_{2a},\xi_{2})$ only have physical significance since they correspond to the locations of $\oa$, namely $(t_{1},0)$ and $(t_{2},0)$ respectively.

Thus, if $\oa$ and $\ob$ are to be in a physically realisable state of relative motion, it is necessary that coordinates $(t,x)\in \ka$ have counterparts $(\tau,\xi)\in \kb$, and as such it is necessary that there exist maps
\[
\begin{aligned}
& \vb{X}:\kb\to\ka  & \quad &\vb{\Xi}:\ka\to\kb \\
& \vb{X}:\mqty(\tau\\ \xi) \mapsto \mqty(t(\tau,\xi)\\ x(\tau,\xi)) & \quad &\vb{\Xi}:\mqty(t\\ x) \mapsto \mqty(\tau(t,x)\\ \xi(t,x)).
\end{aligned}
\]
Moreover, we may translate the coordinates in $\ka$ say, according to $(t,x)\to(t+t_{0},x+x_{0})$ where $t_{0}$ and $x_{0}$ are arbitrary constants, without changing the overall configuration of the observers. In addition, the energy-momentum of the system of observers, as observed from the perspective of $\oa$ is due to the relative motion of $\ob$ essentially (of course there are rest-energies to be associated with each observer), which is characterised by the ratio of \emph{coordinate differences} $\beta c=\frac{x_{2}-x_{1}}{t_{2}-t_{1}}$ only. An arbitrary translation of coordinates as described above is irrelevant to this ratio, and as such our only concern is how the coordinate maps $\vb{X}$ and $\vb{\Xi}$ transform coordinate separations observed in each reference frame.

In principle, we may consider the relative motion of $\oa$ and $\ob$ over arbitrarily small space time intervals, separated by coordinate differences
$$(\dd{\tau},\dd{\xi}) = (\tau+\dd\tau,\xi+\dd\xi)-(\tau,\xi)$$
for any $(\tau,\xi)\in\kb$. The coordinate separations $(\dd \tau,\dd \xi)$ are related to their counterparts $(\dd{t},\dd{x})$ according to
$$
\begin{bmatrix}
\dd{\tau} \\ \dd{\xi}
\end{bmatrix}
=
\begin{bmatrix}
\tau_{t} & \tau_{x} \\ \xi_{t} & \xi_{x}
\end{bmatrix}
\begin{bmatrix}
\dd{t} \\ \dd{x}
\end{bmatrix}
\qquad
\begin{bmatrix}
\dd{t}  \\ \dd{x}
\end{bmatrix}
=
\begin{bmatrix}
t_{\tau} & t_{\xi} \\ x_{\tau} & x_{\xi}
\end{bmatrix}
\begin{bmatrix}
\dd{\tau} \\ \dd{\xi}
\end{bmatrix},
$$
where sub-scripts denote differentiation with respect to the relevant variable.

To deduce the form of these Jacobian matrices, we note that points along the trajectory of $\oa$ in $\ka$ are separated by tangent vectors of the form $(\dd t,0)$, while the trajectory of $\ob$ has tangent vector of the form $(\dd t,\beta c\dd t)$ with reference to $\ka$. Conversely, the trajectory of $\ob$ has a tangent vector $(\dd\tau,0)$ while the tangent vector of $\oa$ is of the form $(\dd\tau,-\beta c \dd\tau)$ with reference to $\kb$ (cf. Figure \ref{fig:motionab}). As such, this means the Jacobian matrices are required to satisfy
\begin{subequations}
\begin{align}
\begin{bmatrix}
\dd{\tau} \\ -\beta c \dd{\tau}
\end{bmatrix}
=
\begin{bmatrix}
\tau_{t} & \tau_{x} \\ \xi_{t} & \xi_{x}
\end{bmatrix}
\begin{bmatrix}
\dd{t} \\ 0
\end{bmatrix} \implies \xi_{t}=-\beta c\tau_{t} \label{eq:Ja}
\\
\begin{bmatrix}
\dd{\tau} \\ 0
\end{bmatrix}
=
\begin{bmatrix}
\tau_{t} & \tau_{x} \\ \xi_{t} & \xi_{x}
\end{bmatrix}
\begin{bmatrix}
\dd{t} \\ \beta c \dd{t}
\end{bmatrix}\implies\xi_{t}=-\beta c\xi_{x},\label{eq:Jb}
\end{align}
from which it follows $\xi_{x}=\tau_{t}$.
To ensure consistency with the special theory of relativity, it is required that a tangent vectors of the form $(\dd{t}, c\dd{t})$ has a counterpart $(\dd{\tau},c\dd{\tau})$, that is to say the speed of light is the same in $\ka$ and $\kb$. As such, it is necessary that
\begin{equation}\label{eq:Jc}
\begin{bmatrix}
\dd{\tau} \\ c\dd\tau
\end{bmatrix}
=
\begin{bmatrix}
\tau_{t} & \tau_{x} \\ \xi_{t} & \tau_{t}
\end{bmatrix}
\begin{bmatrix}
\dd{t} \\ c \dd{t}
\end{bmatrix}\implies\xi_{t}= c^2\tau_{x},
\end{equation}
\end{subequations}
Similar considerations ensure the components of the inverse Jacobian matrix are related according to
$$
x_{\xi} = t_{\tau}\quad t_{\xi}=\frac{1}{c^2}x_{\tau}.
$$
Thus the Jacobian matrix of the coordinate map $\vb{\Xi}:\ka\to\kb$ and its inverse $\vb{X}:\kb\to\ka$ may be written according to
\begin{equation}\label{eq:MatrixLorentz}
\pdv{(\tau,\xi)}{(x,t)}
=
\mqty[\tau_{t} & \tau_{x} \\ c^2\tau_{x} & \tau_{t}]
\iff
\pdv{(t,x)}{(\tau,\xi)}
=
\mqty[t_{\tau} & \frac{1}{c^2}x_{\tau} \\ x_{\tau} & t_{\tau}].
\end{equation}
In addition it is required that the Jacobian of each coordinate map should satisfy
\begin{equation}\label{eq:J}
J=\tau_{t}^2-c^2\tau_{x}^2 = t_{\tau}^2-\frac{1}{c^2}x_{\tau}^2 = 1,
\end{equation}
which simply follows from the symmetry requirement that the Jacobian matrix of $\vb{\Xi}$ is obtained from that of $\vb{X}$ under the change of sign $\beta c\to-\beta c$. More importantly however, it means the maps $\vb{\Xi}:\kb\to\ka$ and $\vb{X}:\ka\to\kb$ are \emph{invertible}, meaning the observers $\oa$ and $\ob$ have mutually consistent trajectories in each reference frame (cf. Figure \ref{fig:motionab}).

\subsection{The Hamilton-Jacobi Equations}
It is clear from \eqref{eq:MatrixLorentz} that the Jacobian matrix of $\vb{\Xi}:\ka\to\kb$ may be characterised by the single function $\tau(t,x)$ only, since the conditions $\tau_{t}=\xi_{x}$ and $\xi_{t}=c^2\tau_{x}$ ensure $\xi(t,x)$ may be obtained from $\tau(t,x)$ essentially by interchanging variables. Meanwhile, the Jacobian matrix of the inverse map $\vb{X}:\kb\to\ka$ is parameterised by the variable $\tau$ only, which is simply a reflection of the fact we do not consider the energy of this system of observers to be dependent on their actual separation. Our goal now is to understand how these maps behave as functions of their respective variables, thereby obtaining a system of partial differential equations (the governing equations) which these maps must satisfy.

\begin{figure}[h]
\centering
\includegraphics[width=\textwidth]{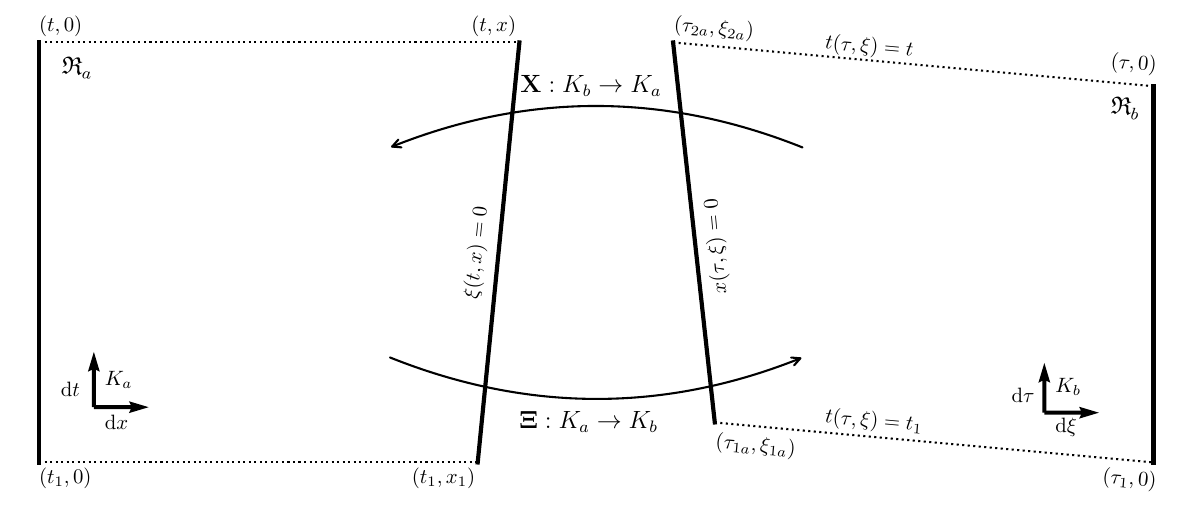}
\caption{The space-time regions $\mathfrak{R}_{a}$ and $\mathfrak{R}_{b}$, induced by the relative motion of $\oa$ and $\ob$, as observed from $\ka$ and $\kb$. The solid black lines are the space-time trajectories of $\oa$ and $\ob$ in each frame, while the dashed lines complete the boundary of the regions $\mathfrak{R}_{a}$ and $\mathfrak{R}_{b}$}\label{fig:areas}
\end{figure}

Figure \ref{fig:areas} shows the space-time regions $\mathfrak{R}_{a}$ and $\mathfrak{R}_{b}$ induced by the relative motion of $\oa$ and $\ob$ as seen from the reference frames $\ka$ and $\kb$.
The action for the coordinate map  $\vb{X}:\kb\to\ka$  is given by
\begin{equation}\label{eq:Sb}
S[\vb{X}]=\frac{m}{2}\int_{\tau_{1}}^{\tau}\vb{X}_{\tau}.\vb{X}_{\tau}\dd{\tau} = \int_{\tau_{1}}^{\tau}L[\vb{X},\vb{X}_{\tau}]\dd{\tau}.
\end{equation}
The notation means $\vb{X}(\tau)\equiv \left(ct(\tau,0),x(\tau,0)\right)\in K_{a}$ which is the image of the map $\mathbf{X}:\kb\to\ka$ applied to the path $\left(\tau,0\right)\in K_{b}$. The constant $E_{0}=mc^2$ is the rest-energy of the observer $\ob$. The inner-product is given by
$$
\vb{X}_{\tau}.\vb{X}_{\tau} = c^2t_{\tau}^2-x_{\tau}^2 = c^2J
$$
where $J$ is the Jacobian of the coordinate map $\mathbf{X}:\kb\to\ka$ (cf. equation \eqref{eq:J}). The constraint $J=1$ is interpreted as a weak equation, to be applied \emph{after} variational derivatives are calculated, in line with the terminology of Dirac (see \cite{Dir1964}). This is of course one form of the action functional governing the relativistic motion of the observer $\ob$, with reference to the frame $\ka$ (see \cite{Gol2002} for further discussion of the various forms this functional may assume).

The associated Hamilton principal function $S[\vb{X}_{1},\vb{X}]$ (see \cite{HF1998} for instance) is obtained by fixing the initial coordinates
$$\vb{X}_{1}=(ct(\tau_{1},0),x(\tau_{1},0))=(ct_{1},x_{1}).$$
Under a variation of the form
$$\vb{X}(\tau)\to \vb{X}(\tau) + \epsilon \uu(\tau),\quad \uu=(u(\tau),w(\tau)),$$
where $\epsilon$ is a constant parameter, Hamilton's principle is simply the requirement
$$\fdv{S}{\vb{X}}:=\eval{\frac{d}{d\epsilon}S[\vb{X}_{1},\vb{X}+\epsilon\uu]}_{\epsilon=0} = 0.$$
This may be written for a general Lagrangian $L[\vb{X},\vb{X}_{\tau}]$ according to
\begin{equation}\label{eq:dS}
\int_{\tau_{1}}^{\tau}\left[\frac{\partial L}{\partial \vb{X}} -\dv{\tau}\frac{\partial L}{\partial \vb{X}_{\tau}} \right].\uu \dd{\tau} + \int_{\tau_{1}}^{\tau}\dv{\tau}\left(\pdv{L}{\vb{X}_{\tau}}.\uu\right)\dd{\tau} = 0
\end{equation}
after integration by parts. Imposing the boundary conditions
$$\uu(\tau_{1}) = \uu(\tau) = (0,0)$$
to an otherwise arbitrary variation $\uu(\tau)$ ensures the first integral on the left-hand side of \eqref{eq:dS} must vanish, thereby yielding the Euler-Lagrange equations
\begin{equation}\label{eq:EL}
\pdv{L}{\vb{X}} -\dv{\tau}\frac{\partial L}{\partial \vb{X}_{\tau}} = (0,0).
\end{equation}
In the specific case
$$L[\vb{X},\vb{X}_{\tau}]=\frac{E_{0}}{2}\left(t_{\tau}^2-\frac{x_{\tau}^2}{c^2}\right),$$
it is clear that $\pdv{L}{\vb{X}}=(0,0)$ while
\begin{equation}
\pdv{L}{t_{\tau}}=E_{0}t_{\tau} \quad \pdv{L}{x_{\tau}} = -\frac{E_{0}x_{\tau}}{c^2}
\end{equation}
and as such
$$\dv{\tau}\left(\pdv{L}{t_{\tau}}, \pdv{L}{x_{\tau}}\right)=(0,0)\implies \dv{\tau}\left(t_\tau,x_\tau\right)=(0,0).$$
That is to say, the Euler-Lagrange equations for the coordinate map $\mathbf{X}:\kb\to\ka$ are  $\dv[2]{\tau}\mathbf{X}(\tau,0) = (0,0)$.

The Hamilton-Jacobi equation follow from the condition $\vb{X}(\tau,0)$ is a physical coordinate map  (i.e. satisfying \eqref{eq:EL}), while the variation is now required to satisfy
$$\uu(\tau_{1})=(0,0) \text{ only},$$
while $\uu(\tau)$ may be arbitrarily chosen. The variation of the action under this perturbation is obtained from \eqref{eq:dS}
\begin{equation}\label{eq:StSx}
\lim_{\epsilon\to0}\frac{S[\vb{X}_{1},\vb{X}+\epsilon\uu] - S[\vb{X}_{1},\vb{X}]}{\epsilon \uu}=\frac{\partial S}{\partial \vb{X}}.
\end{equation}
Expanding the left-hand side explicitly, one obtains
\begin{multline}
\lim_{\epsilon\to0}\frac{S[\vb{X}_{1},\vb{X}+\epsilon\uu] - S[\vb{X}_{1},\vb{X}]}{\epsilon \uu(\tau)}\\
=\frac{1}{\uu(\tau)}\int_{\tau_{1}}^{\tau}\lim_{\epsilon\to0}\frac{L[\vb{X}+\epsilon\vb{u}],\vb{X}_{\tau}+\epsilon\vb{u}_{\tau}]-L[\vb{X},\vb{X}_{\tau}]}{\epsilon}\dd{\tau}\\
=\frac{1}{\uu(\tau)}\int_{\tau_{1}}^{\tau}\left[\pdv{L}{\vb{X}}.\vb{u}+\pdv{L}{\vb{X}_{\tau}}.\vb{u}_{\tau}\right]\dd{\tau}.
\end{multline}
and so upon integrating by parts, we find
\begin{multline}
\lim_{\epsilon\to0}\frac{S[\vb{X}_{1},\vb{X}+\epsilon\uu] - S[\vb{X}_{1},\vb{X}]}{\epsilon \uu(\tau)}\\
=\frac{1}{\uu(\tau)}\int_{\tau_{1}}^{\tau}\left[\pdv{L}{\vb{X}}-\dv{\tau}\pdv{L}{\vb{X}_{\tau}}\right].\vb{u}\dd\tau + \frac{1}{\uu(\tau)}\int_{\tau_{1}}^{\tau}\dv{\tau}\left(\pdv{L}{\vb{X}}.\vb{u}\right)\dd{\tau}.
\end{multline}
Since $\vb{X}(\tau,0)$ is a physical coordinate map it satisfies the Euler-Lagrange equations \eqref{eq:EL}, while the boundary condition $\vb{u}(\tau_{1})=(0,0)$ now means
\begin{equation}
\lim_{\epsilon\to0}\frac{S[\vb{X}_{1},\vb{X}+\epsilon\uu] - S[\vb{X}_{1},\vb{X}]}{\epsilon \uu(\tau)} = \frac{1}{\vb{u}(\tau)}\pdv{L[\vb{X},\vb{X}_{\tau}]}{\vb{X}}.\vb{u}.
\end{equation}
Specifically choosing $\vb{u}(\tau)=(u(\tau),0)$ and $\vb{u}(\tau)=(0,w(\tau))$ separately, it follows at once that
\begin{subequations}
\begin{align}
E &:=\pdv{S}{t} = \pdv{L[\vb{X},\vb{X}_{\tau}]}{t_{\tau}} = E_{0}t_{\tau} \label{eq:canE}\\
p &:=-\pdv{S}{x} = -\pdv{L[\vb{X},\vb{X}_{\tau}]}{x_{\tau}} = \frac{E_{0}}{c^2}x_{\tau}\label{eq:canp},
\end{align}
\end{subequations}
which is the Hamilton-Jacobi equation defining the canonical energy-momentum $(E,p)$ for a general Lagrangian $L[\vb{X},\vb{X}_{\tau}]$, (see \cite{LL2000} for instance). Upon imposing the constraint $J=t_{\tau}^2-\frac{x_{\tau}^2}{c^2}=1$, it follows from \eqref{eq:canE}--\eqref{eq:canp} that
\begin{equation}\label{eq:ge1}
\left(\frac{\partial S}{\partial t}\right)^2 - c^2\left(\frac{\partial S}{\partial x}\right)^2 = E_{0}^2.
\end{equation}
This in turn is equivalent to the relativistic energy-momentum constraint
$$E^2=p^2c^2+E_{0}^2,$$
and for brevity we will use the notation $E_{p}=\sqrt{p^2c^2+E_0^2}$ in what follows.

The Hamiltonian function for the coordinate  map $\mathbf{X}:\kb\to\ka$  is defined as
$$
H=\pp.\vb{X}_{\tau} - L[\vb{X},\vb{X}_{\tau}],\quad  \pp=\left(\pdv{L}{t_{\tau}},-\pdv{L}{x_{\tau}}\right).
$$
This Hamiltonian is manifestly conserved given the canonical energy-momentum defined in \eqref{eq:canE}--\eqref{eq:canp}. In the specific case $L=\frac{E_0}{2}\left(t_{\tau}^2-\frac{x_{\tau}^2}{c^2}\right)$, we have $\vb{P}=\left(E_{0}t_{\tau},\frac{E_{0}}{c^2}x_{\tau}\right)$, while the Hamiltonian is explicitly given by
\begin{equation}
H[\pp,\vb{X}] = E_{0}\left(t_{\tau},\frac{x_{\tau}}{c^2}\right).\left(t_{\tau},x_{\tau}\right)-\frac{E_{0}}{2}\left(t_{\tau}^2-\frac{x_{\tau}^2}{c^2}\right)=\frac{E_{0}}{2}.
\end{equation}
We note that this is not the physical energy of $\ob$ with reference to $\ka$, rather, it is the Hamiltonian function $H[\pp,\vb{X}]$ (see \cite{Gol2002} for instance), which may be used to re-cast \eqref{eq:EL} according to
$$
\dv{\vb{X}}{\tau}=\pdv{H}{\pp},\quad \dv{\pp}{\tau} = -\pdv{H}{\vb{X}},
$$
which are the equations of motion in symplectic form. The second governing equation we require, namely
continuity of energy-momentum in the form $\frac{1}{c^2}\partial_{t} E + \partial_{x} p=0$ or equivalently
\begin{equation}\label{eq:ge2}
\frac{1}{c^2}\pdv[2]{S}{t}-\pdv[2]{S}{x} = 0,
\end{equation}
is consistent with the constraint \eqref{eq:ge1}, since $\partial_{t}\frac{\partial S}{\partial t} = \partial_{t}\frac{\partial L}{\partial t_{\tau}} = \frac{\partial^2L}{\partial t\partial t_{\tau}} = 0$ and likewise for $\frac{\partial^2S}{\partial x^2}$.

Upon using the definitions \eqref{eq:canE}--\eqref{eq:canp} and the constraint $J=1$, we also find
\begin{equation}\label{eq:S[tau]}
\begin{aligned}
\frac{dS}{d\tau} &= \frac{\partial S}{\partial t}t_{\tau} + \frac{\partial S}{\partial x} x_{\tau} \\
                 &= E_{0}t_{\tau}^2-E_{0}\frac{x_{\tau}^2}{c^2}\\
                 &= E_{0},
\end{aligned}
\end{equation}
which is independent of $\tau$. Integrating with respect to $\tau$ we find
\begin{equation}\label{eq:Stau}
\begin{aligned}
\int_{\tau_{1}}^{\tau}\dv{S}{\tau}\dd\tau &= \int_{\tau_{1}}^{\tau}E_{0}\dd\tau \\
                                          &= E_{0}(\tau-\tau_{1}).
\end{aligned}
\end{equation}
That is to say, the coordinate map $\tau(\vb{X})$ is equivalent to the action $S[\vb{X}_{1},\vb{X}]$ with
$$S[\vb{X}_{1},\vb{X}] = E_{0}(\tau(\vb{X})-\tau(\vb{X}_{1})),$$
and in particular, $E_0\tau(t,x)$ must satisfy the system \eqref{eq:ge1} and \eqref{eq:ge2}. Thus, the coordinate map $\bm{\Xi}:\ka\to\kb$ may be characterised entirely in terms of $\tau(t,x)$, which in turn must satisfy the governing equations
\begin{subequations}
\begin{align}
(\partial_{t}\tau)^2 - c^2(\partial_{x}\tau)^2 &=  1 \label{eq:G1}\\
\partial_{t}^{2}\tau-c^2\partial_{x}^2\tau &= 0  \label{eq:G2}.
\end{align}
\end{subequations}
As such, the motion of $\ob$ may be entirely described by the coordinate map $\vb{\Xi}:\ka\to\kb$ or its inverse $\vb{X}:\kb\to\ka$ as illustrated in Figure \ref{fig:areas}.

\begin{remark}
The description of dynamics as flows through appropriate spaces of mappings has a long history, beginning with the celebrated work of Euler \cite{Eul1765} which described the motion of a rigid body as a geodesic flow on the Lie group of rotations of the ambient 3-dimensional space (see \cite{Kol2004} for a contemporary exposition). Euler's approach was extended by Arnold \cite{Arn1966} who recast the equation of motion of a  fluid as a geodesic flow on the group of volume-preserving diffeomorphisms  of the fluid domain. The analysis of changing continuum systems as geodesic flows on appropriate diffeomorphism groups has since been a highly fruitful domain of mathematical physics, see for instance \cite{CK2002, EIK2011, EHKL2016} among many others. Of course there are fundamental differences in the current context, since there is no ambient space to begin with and secondly, the time variable is no longer simply a parameter for the group elements, but a variable actively effected by the coordinate map.
\end{remark}

\section{Coordinate maps and their representations}\label{sec:solutions}
The Hamilton-Jacobi equations for a relativistic point particle in 1+3-dimensions may be written according to
\begin{equation*}
S =  -Et+ \bm{p\cdot x} \quad E= -\frac{\partial S}{\partial t} \quad \bm{p}=\grad S,
\end{equation*}
and were used in the work of Motz \& Selzer \cite{MS1964} to show that the  conservation equation
\begin{equation*}
\frac{1}{c^2}\partial_{t}E + \div\bm{p}=0
\end{equation*}
and the constraint
\begin{equation*}
E^2-\bm{p}^2c^2=m^2c^4
\end{equation*}
yield the  Klein-Gordon equation
\begin{equation*}\label{eq:KG3d}
\frac{1}{c^2}\partial_{t}^2\psi - \laplacian\psi + \frac{m^2c^2}{\hbar^2}\psi = 0,
\end{equation*}
for the \emph{postulated} wave-form
\begin{equation*}
\psi(t,\mathbf{x}) = e^{\frac{i}{\hbar}S}.
\end{equation*}
The point emphasised by Motz \& Selzer is that this ``wave-function'' is associated with a classical particle following a well defined trajectory $(t,\bm{x}(t))$ in (presumably absolute) space-time. As such, even in the classical regime one may associate a wave-form with such a particle, at least in an algebraic sense.

In contrast, our aim in this section is to deduce the existence of such a wave-function from the system of governing equations \eqref{eq:G1}--\eqref{eq:G2}. It will be shown that this wave-form is simply a representation of the map $\tau(t,x)$ and by extension $\bm{\Xi}:\ka\to\kb$,  which in turn is essential to the description of the relative motion of the observers $\oa$ and $\ob$, in the context of relational space and time.

\subsection{Linearity of the coordinate maps}\label{sec:linear}
The main result of this section is that the system \eqref{eq:G1}--\eqref{eq:G2} only admits solutions $\tau(t,x)$ which are linear in $t$ and $x$.

Without imposing assumptions or restrictions, we consider a general solution of the form
\begin{equation}\label{eq:Spsi}
 \tau(t,x) = \Theta(\psi(t,x)),
\end{equation}
with $\psi(t,x)$ being a \emph{representation} of $\tau(t,x)$. Substituting \eqref{eq:Spsi} into the governing equations \eqref{eq:G1}--\eqref{eq:G2} yields
\begin{subequations}
\begin{align}
\left[\left(\partial_{t}\psi\right)^2 - c^2\left(\partial_{x}\psi\right)^2\right]\Theta'(\psi)^2 &=1, \label{eq:psit^2} \\
\left[\partial_{t}^{2}\psi - c^2\partial^2_{x}\psi\right]\Theta'(\psi) + \left[\left(\partial_{t}\psi\right)^2 - c^2\left(\partial_{x}\psi\right)^2\right]\Theta''(\psi) &= 0, \label{eq:psitt}
\end{align}
\end{subequations}
respectively, where $\Theta'(\psi)=\dv{\Theta}{\psi}$.

Equation \eqref{eq:psit^2} applied to equation \eqref{eq:psitt} now yields
\begin{equation}\label{eq:tau''}
\partial^2_{t}\psi - c^2\partial^2_{x}\psi + \frac{\Theta''(\psi)}{\Theta'(\psi)^3} = 0.
\end{equation}
Multiplying \eqref{eq:tau''} by $\partial_{t} \psi$ and using $\partial_{t}\left(\frac{1}{2\Theta'(\psi)^2}\right)=-\frac{\Theta''(\psi)}{\Theta'(\psi)^{3}}\partial_{t}\psi$ it now follows that
\begin{equation}
\frac{1}{2}\partial_{t}\left[(\partial_{t}\psi)^2- \frac{1}{\Theta'(\psi)^2}\right] -  c^2\partial^2_{x}\psi\partial_{t}\psi = 0,
\end{equation}
while substituting $\frac{1}{\Theta'(\psi)^2}=\left(\partial_{t}\psi\right)^2 - c^2\left(\partial_{x}\psi\right)^2$ from equation \eqref{eq:psit^2} we deduce
$$\partial_{x}\psi\partial_{x} \partial_{t}\psi - \partial_{t}\psi\partial_{x}^2\psi = 0.$$
Equivalently, this may be written according to
$$
\partial_{x}\left(\frac{\partial_{x}\psi}{\partial_{t}\psi}\right) = 0.
$$
Multiplying equation \eqref{eq:tau''} by $\partial_{x}\psi$ we also deduce
$$\partial_{t}\left(\frac{\partial_{x}\psi}{\partial_{t}\psi}\right) = 0,$$
and as such $\frac{\partial_{x}\psi}{\partial_{t}\psi}$ is simply a constant.

This means the functions $\partial_{t}\psi$ and $\partial_{x}\psi$ are linearly dependent. It follows that $\psi$ may be written according to
$$
 \psi(t,x) = \phi(\omega t - kx) \iff \frac{\partial_{x}\psi}{\partial_{t}\psi} = -\frac{k}{\omega},
$$
where $\phi(\cdot)$ is yet to be determined, while $\omega$ and $k$ are constants. We denote $s=\omega t-kx$ for brevity, and let $\dot{\phi}(s)=\dv{\phi}{s}$, from which we obtain
\begin{equation}\label{eq:phtphx}
\psi_{t}=\omega\dot{\phi}(s)\quad\psi_{x}=-k\dot{\phi}(s).
\end{equation}
The constraint \eqref{eq:G1} or equivalently \eqref{eq:psit^2} now requires
\begin{equation}\label{eq:phidot}
\left(\omega_{0}\dot{\phi}(s)\Theta'(\phi)\right)^2 = 1,
\end{equation}
where we introduce $\omega_{0}^2 := \omega^2-c^2k^2>0$.
Taking the square-root of \eqref{eq:phidot} we now have $\pm\omega_{0}\dv{\phi}{s}\dv{\Theta}{\phi}= 1$ and so integrating it follows that $\Theta(\phi(s)) = \pm\frac{s}{\omega_{0}}$, or equivalently
\begin{equation}\label{eq:tau}
\tau(t,x) = \Theta(\psi(t,x)) = \pm\frac{\omega t-kx}{\omega_{0}},
\end{equation}
up to an additive constant. Formally, we have applied the inverse function theorem to equation \eqref{eq:phidot} which ensures $\pm\omega_{0}\Theta(\cdot) = \phi^{-1}(\cdot)$  (see \cite{Rud1976} for instance).

Moreover, since $S[t,x]=E_{0}\tau(t,x)$ (cf. \eqref{eq:Stau}), the canonical energy-momentum given by \eqref{eq:canE}--\eqref{eq:canp} now ensure  that
\begin{equation}\label{eq:omegaE}
\begin{cases}
\pdv{S}{t} = E\implies  E_{0}\pdv{t}\left(\frac{\omega t-kx}{\omega_{0}}\right)=E  \implies \frac{E}{E_{0}}=\frac{\omega}{\omega_{0}} \\
-\pdv{S}{x} = p\implies  -E_{0}\pdv{x}\left(\frac{\omega t-kx}{\omega_{0}}\right)=p \implies  \frac{p}{E_{0}}=\frac{k}{\omega_{0}}.
\end{cases}
\end{equation}
for the ``positive energy'' solution $\tau(t,x)=\frac{\omega t-kx}{\omega_0}$.

\subsection{Representations of the coordinate map}\label{sec:reps}
Our aim now is to demonstrate that $\tau(t,x)$ as a solution of \eqref{eq:G1}--\eqref{eq:G2} may be represented as an exponential function of $t$ and $x$ (in line with \cite{MS1964}).

Equation \eqref{eq:psit^2} interpreted as a functional equation for $\Theta(\cdot)$ means that under the re-scaling $\psi\to r\psi$ for a non-zero constant $r$, this equation re-scales as
\[r^2\Theta'(r\psi)^2\left[\psi_{t}^2-c^2\psi_{x}^2\right] = 1 =\Theta'(\psi)^2\left[\psi_{t}^2-c^2\psi_{x}^2\right].\]
Given that both expressions are non-zero, it follows that $\psi_{t}^2-c^2\psi_{x}^2\neq0$ also, and so one may divide by this factor to obtain
\begin{equation}
r^2\Theta'(r\psi)^2 = \Theta'(\psi)^2.
\end{equation}
It is now clear that $r\Theta'(r\psi)$ is independent of $r$, meaning $\Theta'(r\psi)\propto\frac{1}{r\psi}$, thus ensuring $r\Theta'(r\psi)\propto r\frac{1}{r\psi}=\frac{1}{\psi}$. As such, we deduce that $\Theta(\cdot)$ is a logarithmic function of its argument, namely
$$
\Theta(\psi(t,x)) \propto \ln\psi(t,x).
$$
Furthermore, since $\Theta(\psi(t,x))=\tau(t,x)$ by definition (cf. equation \eqref{eq:Spsi}), while $S[t,x]=E_{0}\tau(t,x)$, we may infer
\begin{equation}\label{eq:log}
E_{0}\tau(t,x) = \pm E_{0}\frac{\omega t - kx}{\omega_{0}} = \alpha\ln\psi(t,x),
\end{equation}
where $\alpha$ is a constant action parameter. The function $\psi(t,x)$ is a \emph{representation} of the coordinate map $\tau(t,x)$, and may be explicitly written as
\begin{equation}\label{eq:psi1}
\psi(t,x) = e^{\pm\frac{1}{\alpha}(E t - px)},
\end{equation}
having used equation \eqref{eq:omegaE} to re-write the ratios $\frac{E_{0}\omega}{\omega_{0}}=E$ and $\frac{E_{0}k}{\omega_{0}}=p$.

The other possible solution of \eqref{eq:psitt} arises if we let $\Theta''(\psi)=0$, in which case we also obtain
$$
\psi_{tt}=0\quad \psi_{xx}=0.
$$
This means the functions $\Theta(\cdot)$ and $\psi(\cdot,\cdot)$ are \emph{linear functions} of their respective arguments, and as such
we have
$$
\Theta(\psi(t,x))=\kappa\psi(t,x)
$$
for an arbitrary constant $\kappa$. Meanwhile the inverse function theorem (see \eqref{eq:tau}) now requires
$$
\psi(t,x)=\pm\frac{\omega t-kx}{\kappa\omega_0},
$$
whereupon their composition yields
\begin{equation}\label{eq:psi2}
\tau(t,x)=\Theta(\psi(t,x)) =\kappa\left(\pm\frac{\omega t-kx}{\kappa\omega_0}\right) = \pm\frac{\omega t-kx}{\omega_0}.
\end{equation}

\subsection{Momentum measurement \& de Broglie waves}\label{sec:measure}
In \S\S \ref{sec:linear}--\ref{sec:reps} it has been shown that the coordinate map $\tau(t,x)$ governed by \eqref{eq:G1}--\eqref{eq:G2} is necessarily linear with $\tau(t,x)=\pm\frac{Et-px}{E_{0}}$ and has a representation of the form $E_{0}\tau(t,x)=\alpha\ln\psi(t,x)$. Combining these observations then it was found necessary that the representation $\psi(t,x)$ is of the form
$$
\psi(t,x) = \exp{\pm\frac{1}{\alpha}(Et-kx)}.
$$
It is already clear $\alpha$ must have units $[\alpha] = $ \SI{}{\joule\second} and so the choice $\hbar$ is obvious. To ensure the representation $\psi(t,x)$ corresponds to a de Broglie wave of the form \eqref{eq:deB1}, it is also necessary to show $\alpha$ is imaginary, which is the aim of the current section.

\begin{figure}[h!]
\centering
\captionsetup{width=\textwidth}
\includegraphics[width=\textwidth]{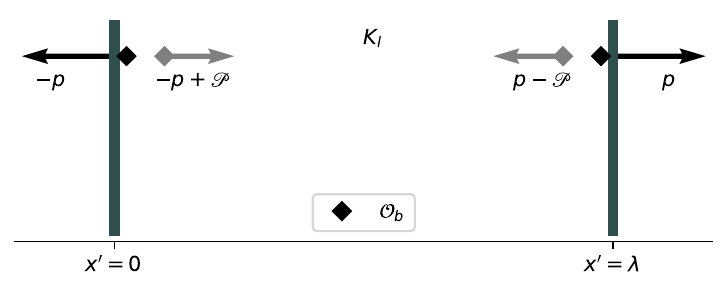}
\caption{The measurement of $\ob$'s momentum by collision with massive plates of equal rest-energy $\eb_{0}$.}\label{fig:measure}
\end{figure}
Figure \ref{fig:measure} shows a very simple apparatus consisting of two massive plates $\pl$ and $\pr$, initially at $x'=0$ and $x'=\lambda$ respectively, both initially static with reference to the frame $K_{l}$ (the rest-frame of $\pl$) and each with rest energy $\eb_{0}$. It is supposed the point-like observer $\ob$ is located at some $x'\in(0,\lambda)$, and interacts with either plate only by collision. Upon collision $\ob$ undergoes a change of momentum, thereby imparting  momentum to one of these plates.
Measurement of momentum means $\ob$ impacts one of the plates and sets it in motion relative to the other.  Immediately after impact the plates are again inertial observers, since there is no further interaction to impart momentum to either plate.

Since $K_{l} \owns (t',x')$ is the rest-frame of $\pl$,  then its coordinates with reference to this frame will always be of the form $\left(t',0\right)$; those of $\pr$ will be of the form $(t',\lambda)$ prior to collision. Similarly, $K_{r} \ni (t^*,x^*)$ is the rest-frame of $\pr$ whose coordinates are always of the form $\left(t^*,0\right)$ in this frame; those of $\pl$ are of the form $(t^*,-\lambda)$ initially. Prior to collision it makes sense to identify coordinate frame $K_{l}$ and $K_{r}$ as a single reference frame $K\owns (t,x)$, since both reference frames see the observers $\pl$ and $\pr$ at rest, and so are equivalent up to constant translations.

At the moment of measurement as observed from the frame $K_{l}$, it appears the observer $\pr$ changes energy-momentum according to $(\eb_{0},0)\to (\eb,\mb)$ where $\eb^2=\mb^2c^2+\eb_{0}^2$ and $\mb>0$ is assumed. Meanwhile the momentum of $\ob$ changes according to $(E,p)\to (E_{1},p-\mb)$ (cf. Figure \ref{fig:measure}). Naturally, the energy-momentum of $\pl$ is \emph{always} $(\eb_{0},0)$ in the frame $K_{l}$ while the observer $\ob$ is interpreted to occupy the location $x'=\lambda$ upon collision. Conversely, in the frame $K_{r}$ the observer $\pl$ changes its energy-momentum according to $(\eb_{0},0)\to (\eb,-\mb)$ and the energy-momentum of $\ob$ changes according to $(E,-p)\to (E_{1},-p+\mb)$. In this frame of reference the observer $\ob$ is interpreted to appear at $x^*=-\lambda$ upon impact, and by definition the energy-momentum of $\pr$ is \emph{always} $(\eb_{0},0)$.

Given that $\pl$ and $\pr$ are in uniform relative motion before and after collision with $\ob$, it follows from \S \ref{sec:reps} the coordinate map $\mathbf{X^*}:K_{l}\to K_{r}$ characterised by $t^*(t',x')$ has a representation of the form
$$
\psi(t',x') =
\begin{cases}
e^{\frac{1}{\alpha}\eb_{0}(t'_{0}-t')},\quad t'< t'_{0} \\
e^{\frac{1}{\alpha}\left(\eb (t'-t'_{0})-\mb x'\right)} \quad t'\geq t'_{0}
\end{cases}
$$
where the impact occurs at time $t'_{0}$ with reference to $K_{l}$. Upon impact the proper-time $t^*$ of the observer $\pr$  changes according to
\[\frac{\alpha}{\eb_{0}}\ln e^{\frac{1}{\alpha}\eb_{0}(t'_{0}-t')} \to \frac{\alpha}{\eb_{0}}\ln e^{\frac{1}{\alpha}\left(\eb (t'-t'_{0})-\mb x'\right)},\]
from the perspective of the observer $\pl$. However, according to the observer $\pr$ its own time coordinate is continuous, while it is the time coordinate of $\pl$ which undergoes a corresponding change during collision with $\ob$. Continuity of the $t^*$-coordinate upon impact now requires
\begin{equation}\label{eq:limit}
\lim_{t'\to t'_0}e^{\frac{1}{\alpha}\eb_{0}(t'_{0}-t')} = \lim_{t'\to t'_0}e^{\frac{1}{\alpha}\left(\eb (t'-t'_{0})-\mb \lambda\right)} \iff e^{-\frac{\mb \lambda}{\alpha}} = 1.
\end{equation}
Since $\lambda\neq 0$ and $\mb>0$ by assumption, continuity of $\psi(t,x)$ at $t'_{0}$ is satisfied only when the argument of the exponential is of the form $2 \pi n i$ for $n\in \mathbb{Z}$. Hence, we deduce
\begin{equation}\label{eq:ihbar}
\alpha = i\hbar,\quad \mb = \frac{2\pi n\hbar}{\lambda},
\end{equation}
and so the action parameter $\alpha$ is imaginary as anticipated.

Additionally, if we allow for the possibility that $\ob$ is completely absorbed by the plates during the collision, then it must be the case that the energy-momentum of $\ob$ relative to the plates before collision should have the specific values
$$
\left(E_{n},p_{n}\right) = \left(\sqrt{E_{0}^2+c^2p_{n}^2},\frac{2\pi n\hbar}{\lambda}\right),\quad n\in \ZZ.
$$
Thus the allowable energy-levels of $\ob$ relative to the plates prior to collision are discrete, as opposed to a continuous spectrum one expects in the classical relativistic framework.

With $\alpha=i\hbar$ it is now clear that the coordinate transformation between the rest frames of inertial observers may in general be represented by wave-forms
\begin{equation}\label{eq:de Broglie}
\psi^{(\pm)}(t,x) = e^{\mp\frac{i}{\hbar}\left(\ep t-px\right)},
\end{equation}
where $\psi^{(\pm)}(t,x)$ are called ``positive/negative energy'' solutions respectively, in line with standard terminology (see \cite{Nik2007} for instance).
It has been deduced in equations \eqref{eq:Stau}, \eqref{eq:log} and \eqref{eq:ihbar} that the action $S[t,x]$ may be defined in terms of $\psi(t,x)$ according to $S[t,x]=E_{0}\tau(t,x)=i\hbar\ln\psi(t,x)$. The canonical energy-momentum as given by $E_{p}=\pdv{S}{t}$ and $p=-\pdv{S}{x}$ (cf. \eqref{eq:canE}--\eqref{eq:canp}) may be written according to
\begin{equation}\label{eq:evals}
\begin{aligned}
\ep=\pdv{t}\left(i\hbar\ln\psi\right)=\frac{1}{\psi}i\hbar\p{t}\psi=\bar{\psi}i\hbar\p{t}\psi,\\
p=-\pdv{x}\left(i\hbar\ln\psi\right)=-\frac{1}{\psi}i\hbar\p{x}\psi=-\bar{\psi}i\hbar\p{x}\psi
\end{aligned}
\end{equation}
where $\bar{\psi}=\frac{1}{\psi}$ corresponds to the complex conjugate of $\psi$ specifically when $\psi(t,x)=e^{-\frac{i}{\hbar}(E_{p}t-px)}$ is an energy-momentum eigenfunction. Alternatively, we may re-write these canonical relations as eigenvalue problems of the form
$$
i\hbar\p{t}\psi=E\psi\qquad -i\hbar\p{x}\psi = p\psi
$$
familiar from (non)relativistic quantum mechanics (see \cite{Dir2007, Gre2000} for instance). Both representations $\psi$ and $\bar{\psi}$ satisfy the Klein-Gordon equation
\begin{equation}\label{eq:KG}
\frac{1}{c^2}\pdv[2]{\psi}{t} - \pdv[2]{\psi}{x} + \frac{m^2c^2}{\hbar^2} \psi = 0.
\end{equation}
Thus, de Broglie waves as per Dirac's terminology (see \cite{Dir2007}, p. 120) emerge as a representation of the $\tau$-component of the coordinate map $\bm{\Xi}:\ka\to\kb$, and so represents the trajectory of $\ob$ (i.e. $(\tau,0)\in\kb$) with reference to $\ka$.

\section{Implications and results}

\subsection{Eigenfunctions and superpositions}\label{sec:(E,p)-eigenfunctions}
The $\tau$-representation given in equation \eqref{eq:de Broglie} is an eigenfunction of the linear operators $i\hbar\partial_{t}$ and $-i\hbar\partial_{x}$, whose corresponding eigenvalues are simply the energy-momentum of the observer $\ob$ with reference to the frame $K_{a}$. The nonlinear constraint \eqref{eq:ge1} has a particularly elegant geometric interpretation in the relational context, since one may reformulate the coordinate map \eqref{eq:MatrixLorentz} according to
\begin{equation}\label{eq:Jacobian}
\begin{bmatrix}
\dd\tau\\ \dd\xi
\end{bmatrix} =
\begin{bmatrix}
\tau_{t} & \tau_{x} \\ \xi_{t} & \xi_{x}
\end{bmatrix}
\begin{bmatrix}
\dd t\\\dd x
\end{bmatrix} =
\begin{bmatrix}
\tau_{t} & \tau_{x} \\ c^2\tau_{x} & \tau_{t}
\end{bmatrix}
\begin{bmatrix}
\dd t\\\dd x
\end{bmatrix}
\end{equation}
in which case $\tau_{t}^2-c^2\tau_{x}^2 =1$ is equivalent to
$
\det\begin{bmatrix}\tau_{t} & \tau_{x} \\ \xi_{t} & \xi_{x}\end{bmatrix} =  1.
$

Hence, the Jacobian of the coordinate transformation $\bm{\Xi}:K_{a}\to K_{b}$ is required to be one, thus ensuring this map is invertible. Specifically, it means that a trajectory $(\dd{t},\dd{x})$ in $K_{a}$ has as counterpart $(\dd{\tau},\dd{\xi})$ with reference to $K_{b}$ and vice-versa. In particular it means that a trajectory of $\ob$  in $K_{b}$ whose tangent $(\dd{\tau},0)$ has a counterpart $(\dd{t},\dd{x})$ in $K_{a}$, while simultaneously the trajectory of $\oa$ in $K_{a}$, whose tangent $(\dd{t},0)$,  has a counterpart $(\dd{\tau},\dd{\xi})$ in $K_{b}$, (cf. Figure \ref{fig:motionab}). As such, these observers appear as point-like bodies moving with reference to the rest-frame of their counterpart. This is {only} possible since the conditions \eqref{eq:G1}--\eqref{eq:G2} are both satisfied for the coordinate map $E_{0}\tau(t,x)=i\hbar\ln\psi(t,x)$ when $\psi(t,x)$ is an energy-momentum eigenfunction.

Contrarily, given the linearity of \eqref{eq:KG} it is clear that superpositions of the form
$$\varphi(t,x) = \sum_{n=1}^{N}a_{n}e^{\frac{i}{\hbar}(E_{n}t-p_{n}x)}\quad E_{n}^2=E_0^2+p_n^2c^2$$
are also valid solutions of this wave equation. However, such a superposition cannot represent a physically realisable coordinate map from $K_{a}$ to $K_{b}$ since the non-linear constraint \eqref{eq:G1} is not satisfied for $S=i\hbar\ln\varphi$. This is not to say $\ob$ as described by $\varphi$ becomes somehow de-localised in space-time, It always has a precise location $(\tau,0)\in K_{b}$. Rather, it is the case there is no longer a precise correspondence of the form \eqref{eq:MatrixLorentz} between the frames $K_{a}$ and $K_{b}$, and so the trajectory $(\dd{\tau},0)$ in $\kb$ no longer has a precise counterpart with reference to $\ka$ satisfying all the required axioms of special relativity.

\subsection{Representations as probability amplitudes}\label{sec:probability}
The eigenfunctions
$$
\psi(t,x)=e^{-\frac{i}{\hbar}(E_{p}t-px)}\quad \bar{\psi}(t,x)=e^{\frac{i}{\hbar}(E_{p}t-px)}
$$
are the permissible representations of the coordinate map $\vb{\Xi}:\ka\to\kb$ and its inverse $\vb{X}:\kb\to\ka$ when $\oa$ and $\ob$ are in a well-defined state of uniform relative motion in the context of relational space-time.
It is well established that a solution $\varphi(t,x)$ of the wave-equation \eqref{eq:KG} has an associated conserved current density
\begin{equation}
(\rho(t,x),j(t,x))=\frac{i}{2}\left(\bar{\psi}\psi_{t}-\bar{\psi}_{t}\psi,-c^2(\bar{\psi}\psi_{x}-\bar{\psi}_{x}\psi)\right),
\end{equation}
with the continuity equation
$$
\pdv{\rho}{t}+\pdv{j}{x}=0
$$
and immediate consequence of \eqref{eq:KG}.

The standard argument against interpreting $\psi(t,x)$ as a probability density relies on the fact that $\rho(t,x)$ is not always non-negative, (see \cite{Nik2007} for a comprehensive discussion of this objection). For instance for a positive energy solution $\psi^{(+)}(t,x)=e^{-\frac{i}{\hbar}(E_{p}t-px)}$ one finds
$$
\rho^{(+)}(t,x) = \frac{i}{2}\left(\bar{\psi}^{(+)}\psi^{(+)}_{t}-\bar{\psi}^{(+)}_{t}\psi^{(+)}\right) = \frac{E_{p}}{\hbar},
$$
while for the associated ``negative energy'' solution $\psi^{(-)}(t,x)=e^{\frac{i}{\hbar}(E_{p}t-px)}$ once finds
$$
\rho^{(-)}(t,x) = \frac{i}{2}\left(\bar{\psi}^{(-)}\psi^{(-)}_{t}-\bar{\psi}^{(-)}_{t}\psi^{(-)}\right) = -\frac{E_{p}}{\hbar}.
$$
However, we note from \eqref{eq:tau} that the governing equations \eqref{eq:G1}--\eqref{eq:G2} allow for coordinate maps of the form
$$
\tau(t,x) = \pm\frac{\omega t-kx}{\omega_{0}} = \pm\frac{E_{p}t-kx}{E_{0}},
$$
while equation \eqref{eq:psi1} (with $\alpha=i\hbar$) ensures $\psi^{(\pm)}(t,x)$ are simply the representations of $\tau(t,x) = \pm\frac{E_{p}t-kx}{E_{0}}$.

In the framework of relational space-time this does not imply $\ob$ somehow has a negative energy, the rest energy of $\ob$ is always $E_{0}>0$. Rather, it simply means the the time-like axes of the reference frames $\ka$ and $\kb$ label events in reverse-chronological order. However, given the absence of an ambient absolute space-time, there is no sense in which the time-coordinate of either observer is progressing ``backwards'', since instants of time are only distinguished by the changing relative configurations of the observers. In other words, $t$ and $\tau$ are only progressing in reverse order compared to each other, that is all one can say. An interesting discussion of the statistical (thermal) emergence of a preferred time direction may be found in \S 3.4 of \cite{Rov2004}.

We note that the full current $(\rho^{(-)}(t,x),j^{(-)}(t,x))$ for the representation $\psi^{(-)}(t,x)$ is now given by
\begin{equation}
 (\rho^{(-)}(t,x),j^{(-)}(t,x)) = \left(-\frac{E_{p}}{\hbar},\frac{c^2p}{\hbar}\right).
\end{equation}
Meanwhile, the components of this probability current with reference to $\kb$ are obtained from the Jacobian matrix of the map $\vb{X}:\kb\to\ka$, which itself is the inverse of the Jacobian matrix of $\vb{\Xi}:\ka\to\kb$. The Jacobian matrix obtained from the map $\tau(t,x)=-\frac{E_{p}t-px}{E_{0}}$ is now given by
$$
\pdv{(\tau,\xi)}{(t,x)} = \mqty[\tau_{t} & \tau_{x} \\ c^2\tau_{x} & \tau_{t}] = \frac{1}{E_{0}}\mqty[-E_{p} & p \\ c^2p & -E_{p}],
$$
while its inverse is now given by
\begin{equation}
 \pdv{(t,x)}{(\tau,\xi)} = \mqty[t_{\tau} & \frac{1}{c^2}x_{\tau} \\ x_{\tau} & t_{\tau}] = -\frac{1}{E_{0}}\mqty[E_{p} & p \\ c^2p & E_{p}].
\end{equation}
The current-density $(\rho^{(-)},j^{(-)})$ with reference to $\kb$ now takes the form
\begin{equation}
 \mqty[\rho^{(-)}(\tau,\xi) \\ j^{(-)}(\tau,\xi)] = -\frac{1}{E_{0}}\mqty[E_{p} & p \\ c^2p & E_{p}]\mqty[-\frac{E_{p}}{\hbar} \\ \frac{c^2p}{\hbar}] = \mqty[\frac{E_{0}}{\hbar} \\ 0].
\end{equation}
That is to say, the probability current associated with the motion of $\ob$ along its own $\tau$-axis is the constant $\omega_{0}=\frac{E_{0}}{\hbar}>0$, while there is zero probability current along the $\xi$-axis as $\ob$ is always at $\xi=0$ by definition.

In conclusion, $\rho(t,x)=\frac{i}{2}(\bar{\psi}\psi_{t}-\bar{\psi}_{t}\psi)$ is an acceptable probability density associated with $\ob$, since its counterpart in the rest frame of $\ob$, namely $\rho(\tau,\xi)$, is always positive-definite. Negative probability densities associated with $\ob$ arise in other reference frames only because the times axes of the respective reference frames may not have the same orientation as $\kb$. However, in the context of relational space time, there is no absolute notion as to which time-axis is forward pointing. Rather, there are only relative orientations of time-axes in the framework of relational space-time, and so such negative probability densities are not immediately problematic.\footnote{This was alluded to by Kastner in footnote 3 of \cite{Kas2013}}

\subsection{Relational Quantum Mechanics}
A central point of the argument in \S \ref{sec:measure} is that the observers $\pl$ and $\pr$ are both always inertial in their own rest-frames, and the acceleration of the pair upon impact with $\ob$ is only defined in relative terms. This is apparently consistent only in the relational space-time framework. Moreover, the derivation presented here appears to be consistent with Rovelli's Relational Quantum Mechanics (RQM) \cite{Rov1996, Rov2018}, whereby the properties of a system are not absolutes. In particular, the perceived location and momentum of $\ob$ upon impact with the apparatus depends on the frame of reference adopted for the measurement.

Indeed the physical properties of a system, in this case the energy-momentum of $\pl$ and $\pr$, is a characteristic of interaction between the observers, specifically it is a property of the coordinate maps between their respective rest-frames (cf. \cite{LMB2018}). It is also clear the observer $\ob$ does not have an \emph{absolute location} in this experiment, its apparent location is contingent on the frame of reference used for the observation.

In line with the three characteristics of quantum systems proposed by Rovelli in \S 2 of \cite{Rov2018}, namely
\begin{enumerate}[a)]
 \item Discreetness appearing as an intrinsic feature of nature
 \item The probabilistic character of predictions
 \item Physical variables of one system are defined only relative to another physical system
\end{enumerate}
all emerge or are at least accommodated within the relational framework developed here. In \S \ref{sec:measure} it became apparent that the observer $\ob$ could only impart energy-momentum in discrete units to the plates of the apparatus illustrated in Figure \ref{fig:measure}. This in turn ensures the energy-momentum of $\ob$ can only assume discrete values relative to this apparatus, in the framework of relational space-time. In \S \ref{sec:probability} it was argued the the usual objection to $\rho=i(\bar{\psi}\psi_{t}-\bar{\psi}_{t}\psi)$ being an appropriate probability density are not entirely supported in the context of relational space-time. Thus it appears one may naturally associate a probability current with a coordinate map $\vb{\Xi}:\ka\to\kb$ via its representation $\psi(t,x)$, thereby ensuring the associated motion of $\ob$ relative to the frame $\ka$ acquires a probabilistic character. Finally, the motion of observers $\oa$ and $\ob$ has been described in a framework where this motion is only defined in relative terms. In particular the energy-momentum of one observer relative to another is in reality a property of the system of observers, and not intrinsic to either observer in particular.

Thus the description of relative motion presented here appears to lend support to Rovelli's hypothesis (see \cite{Rov2004} pp. 220--221) that the relational character of states in RQM is connected to the relational framework of space and time.

\section{Outlook}
The coordinate map $\tau(t,x)$ and its representation $\psi(t,x)$ have been represented throughout in terms of the dynamical variables $(E_{p},p)$ of the associated observers. Nevertheless, we have been primarily concerned with states of uniform relative motion only, and so our considerations have been largely kinematic. As such, the major issue the current work does not address is the case of bodies in states of relative acceleration in the context of relational space-time, (see \cite{Bar1974} for an interesting  Machian approach to inertial and Newtonian forces).

The energy-momentum of $\ob$ relative to $\oa$ may be endowed with a space-time dependence by the introduction of potential fields $\Phi(t,x)$ and $A(t,x)$, whereby the eigenvalue problems \eqref{eq:evals} are modified according to
\begin{equation}
 \left(i\hbar\p{t}+\Phi(t,x)\right)\psi = \mathcal{E}(t,x)\psi  \qquad \left(-i\hbar\p{x}+A(t,x)\right)\psi = \mathcal{P}(t,x)\psi.
\end{equation}
The energy-momentum of $\ob$ with reference to $\kb$ may be obtained from the canonical energy-momentum $\left(\mathcal{E}(t,x),\mathcal{P}(t,x)\right)$ by the Jacobian matrix of $\bm{\Xi}:\ka\to\kb$ characterised by $\tau(t,x)=\frac{E t-px}{E_{0}}$, where $(E,p)$ are the instantaneous energy-momentum of $\ob$ with reference to $\ka$. The relational framework developed here relies on the assumption there is always a reference frame in which the observer $\ob$ has energy-momentum $(E_{0},0)$ (i.e. the reference frame $\kb$). Likewise, in the frame $\ka$ it is required the observer $\oa$ should always have energy-momentum $(m_{a}c^2,0)$. This in turn may be attained if $(\Phi(t,0),A(t,0))=(0,0)$ (implemented by choice) and $(\Phi(t,x_{b}),A(t,x_{b}))=(0,0)$. This second condition may be achieved in the contexts of the above eigenvalue problems by the introduction of a local phase shift $\psi \to e^{i\Lambda}\psi$, with $(\Phi,A)\to(A+\Lambda_{t},\Phi-\Lambda_{x})$. By an appropriate choice of $\Lambda$ then in principle it should be possible to eliminate the potentials $\Phi$ and $A$ as a specific location.  In this way it should always be possible to ensure the energy-momentum of the observers $\oa$ and $\ob$ have the appropriate values in the respective rest frames $\ka$ and $\kb$.

As a second approach to introducing relative acceleration in this relational framework, in Figure \ref{fig:areas} we observe that the space-time regions $\mathfrak{R}_{a}$ and $\mathfrak{R}_{b}$ have equal areas with
\[\mathcal{A}[\tau]=\iint_{\mathfrak{R}_{a}}\left(\tau_{t}^2-c^2\tau_{x}^2\right)\dd x\dd t = \iint_{\mathfrak{R}_{b}}\dd\xi\dd\tau,\]
where $J=\tau_{t}^2-c^2\tau_{x}^2$ is the Jacobian of $\vb{\Xi}:\ka\to\kb$. If the constraint $J=1$ is imposed as a weak equation again, then $\mathcal{A}[\tau]$ may be interpreted an action functional for the coordinate map $\tau(t,x)$, whose Euler-Lagrange equations are precisely
$$
\frac{1}{c^2}\tau_{tt}-\tau_{xx} = 0\qquad \tau_{t}^2-c^2\tau_{x}^2=1.
$$
If we allow for the generalisation $J=f(t,x)$ as opposed to $J=1$, then it may be possible to consider the relative motion of $\oa$ and $\ob$ in a curved space-time in this relational framework.

A third interesting problem that may be worth investigating in the relational space-time framework arises when we consider the approach adopted by Pauli \& Weisskopf \cite{PW1934} in addressing the negative probability densities associated with the solutions of \eqref{eq:KG}. In this approach, the solutions $\psi(t,x)$ were interpreted as field operators, subject to the equal time commutator relations
$$
\left[\bar{\psi}_{t}(t,x),\psi(t,y)\right]=i\hbar\delta(x-y)\qquad \left[\psi_{t}(t,x),\bar{\psi}(t,y)\right]=i\hbar\delta(x-y),
$$
thereby ensuring the field $\psi(t,x)$ is also endowed with particle-like characteristics.
An interesting question now is whether such commutation relations naturally arise in the relational space-time framework, and if so what are the implications for solutions of the Klein-Gordon equation \eqref{eq:KG}. In particular, given that the commutators apparently preclude the possibility of specifying $\psi(0,x)$ and $\psi_{t}(0,x)$ simultaneously, this leads one to question in what sense solutions of this wave equation exist in the sense of a well-posed Cauchy problem.

\section*{Declarations}
\subsection*{Conflicts of Interest}
The author declares there are no conflicts of interest, financial or otherwise,  related to this work.

\subsection*{Funding}
No funding was obtained for the preparation of this manuscript.

\section*{Acknowledgements}
The author is grateful to the referees for several helpful comments and suggestions.

\providecommand{\noopsort}[1]{}

\end{document}